\documentclass[]{aa}  

\usepackage{graphicx}
\usepackage{xcolor}

\usepackage{txfonts}

\newcommand{\beq}{\begin{equation}}
\newcommand{\eeq}{\end{equation}}
\newcommand{\beqa}{\begin{eqnarray}}
\newcommand{\eeqa}{\end{eqnarray}}

\usepackage{natbib}
\bibpunct{(}{)}{;}{a}{}{,}

\begin{document}

  \title{Magnetohydrodynamic  shallow water equations with the alpha effect: Rossby-dynamo waves in solar--stellar tachoclines}

   \author{T. V. Zaqarashvili
          \inst{1,3,4}
          \and
          M. Dikpati\inst{2}
         \and
          P. A. Gilman \inst{2}      
 }

   \institute{Institute of Physics, University of Graz, Universit\"atsplatz 5, 8010, Graz, Austria\\
              \email{teimuraz.zaqarashvili@uni-graz.at}
              \and
          High Altitude Observatory, NCAR, 3080 Center Green Drive, Boulder, CO 80301, USA
                                          \and
             Department of Astronomy and Astrophysics at Space Research Center, Ilia State University, 
Kakutsa Cholokashvili Ave 3/5, 0179 Tbilisi, Georgia
                      \and
            Evgeni Kharadze Georgian National Astrophysical Observatory, 
Mount Kanobili, 0301 Abastumani, Georgia \\  
             }

   \date{Accepted: }

  \abstract
   {The activity of Sun-like stars is governed by the magnetic field, which is believed to be generated in a thin layer between convective and radiative envelopes.  The dynamo layer, also called the tachocline,  permits the existence of Rossby waves (r-modes) described by magnetohydrodynamic shallow water models, which may lead to short-term cycles in stellar activity.}
   {Convective cells penetrate into the layer creating an overshoot upper part, where they transport an additional energy for vigorous activity. The aim of this paper is to study the influence of overshooting convection on  the dynamics of Rossby waves in the tachoclines of Sun-like stars.}
   {Here we write the magnetohydrodynamic shallow water equations with the  effect of the penetrative convection and study the dynamics of wave modes in the layer.  }
   {The formalism leads to the excitation of new oscillation modes connected with the dynamo coefficient, $\alpha$, causing periodic modulations of all parameters in the tachocline. The modes are coupled with the Rossby waves resulting mutual exchange of convective and rotation energies. The timescales of Rossby-dynamo waves, for certain parameters, correspond to Schwabe ($ \sim$ 11 years) and Rieger ($\sim$ 150-170 days) cycles as observed in solar activity.    }
   {The waves provide a new paradigm for internal magnetism and may drive the dynamos of Sun-like stars. Theoretical properties of  the waves and observations can  be used for magneto-seismological sounding of stellar interiors. }


   \maketitle

\section{Introduction}

Stellar activity has a tremendous influence on the evolution of exoplanets through flares and coronal mass ejections \citep{Lammer2012}. The activity of Sun-like stars changes throughout their evolution, and often has several timescales of quasi-periodic variations.  Long-term studies of our Sun show that solar activity varies over three main timescales. The most important cycle implies  variations over 10--20  years \citep{Schwabe1844}, which is accompanied by longer cycles of $\sim$ 100 years \citep{Gleissberg1939}  and shorter cycles of $\sim$ 150--250 days  \citep{Rieger1984}. Stellar activity also shows multi-scale temporal variations \citep{Saar1999}. Solar and stellar activity is generally explained in terms of dynamo action incorporating rotation and convection, though many uncertainties still remain \citep{Charbonneau2020}. The magnetic fields and cycles of the Sun and Sun-like stars are believed to be generated in a thin layer below the convection zone called the tachocline \citep{Spiegel1992}, which is estimated to be thinner than the local density scale height. As the convection starts above the tachocline, the temperature gradient must be at least slightly superadiabatic in the overlying area. Consequently, the buoyancy force is positive above the layer, and hence its upper surface does not feel any gravitational force from above (similar to the ocean--atmosphere interface on the Earth). Therefore, the shallow water approximation can be used to study the processes with large horizontal spatial scales. Due to the existence of the magnetic field in the layer, a magnetohydrodynamic (MHD) description of the plasma processes is inevitable \citep{Gilman2000}. 

A shallow water description implies Rossby waves, large-scale oscillatory motions that arise as a result of conservation of total (planetary+relative) vorticity together with the latitudinal variation of the Coriolis parameter \citep{Rossby1939}. Rossby waves govern the large-scale dynamics of the Earth's atmosphere and oceans, and recent observations showed their existence on the Sun and stars \citep{VanReeth2016,Lopten2018,Zaqarashvili2021}. Moreover, in the tachocline the Rossby waves are modified by the magnetic field so that total vorticity is no longer conserved due to torques on fluid elements by the Lorentz force. The waves  may lead to the quasi-periodic emergence of magnetic flux toward the surface producing Rieger-type cycles in solar--stellar activity \citep{zaqarashvili2010,Dikpati2017,Breton2024}. A combination of observed periods and Rossby wave theory then may account for the seismologic estimation of the tachocline magnetic field,  which can be of importance to test the different dynamo models \citep{Gurgenashvili2016}.

The downflows of convective cells penetrate into the tachocline due to their inertia and create an upper overshoot layer (Fig. 1). The depth of the penetrative convection is somewhere between 30 \% and 40 \% of the local pressure scale height and can be approximated as $\sim$ 20 Mm \citep{Rempel2004}. The temperature gradient is slightly sub-adiabatic in the region so that $\delta= | \nabla -\nabla_{ad}| \approx 10^{-6}-10^{-4}$, where $\delta$ is the fractional deference between the actual and adiabatic temperature gradients \citep{Dikpati2001a}. The conditions are significantly different in the lower part of the tachocline, where the temperature gradient is believed to be significantly sub-adiabatic. Therefore, the tachocline can be naturally divided into two sublayers: an inner radiative layer and an outer overshoot layer \citep{Schecter2001}. The MHD shallow water equations of \citet{Gilman2000} do not directly include the effects of overshooting. One of the important consequences of the convection is the $\alpha$ effect, which implies the influence of small-scale turbulent motions on the large-scale dynamics of the magnetic field \citep{Parker1993}. \citet{Deluca1986, Deluca1988} introduced the $\alpha$ effect in shallow water MHD systems with certain limitations (see also  \citep{Deluca}). They considered only an axisymmetric case, so that the variation in the toroidal direction was  ignored. Next, they considered only a Cartesian frame so that the rotation axis is always parallel to the vertical coordinate. Therefore, the Rossby waves are neglected in the consideration. Though the approach is  important  to study the kinematic dynamo in the tachocline, the dynamics of Rossby waves and their coupling to dynamo waves are not included. In this article we include the penetrative convection into general MHD shallow water equations as an additional  $\alpha$  term  to provide the coupling of Rossby and dynamo waves leading to the generation of cyclic magnetic fields and the resulting stellar activity.

\begin{figure}[ht!]
\includegraphics[angle=0,width=9cm]{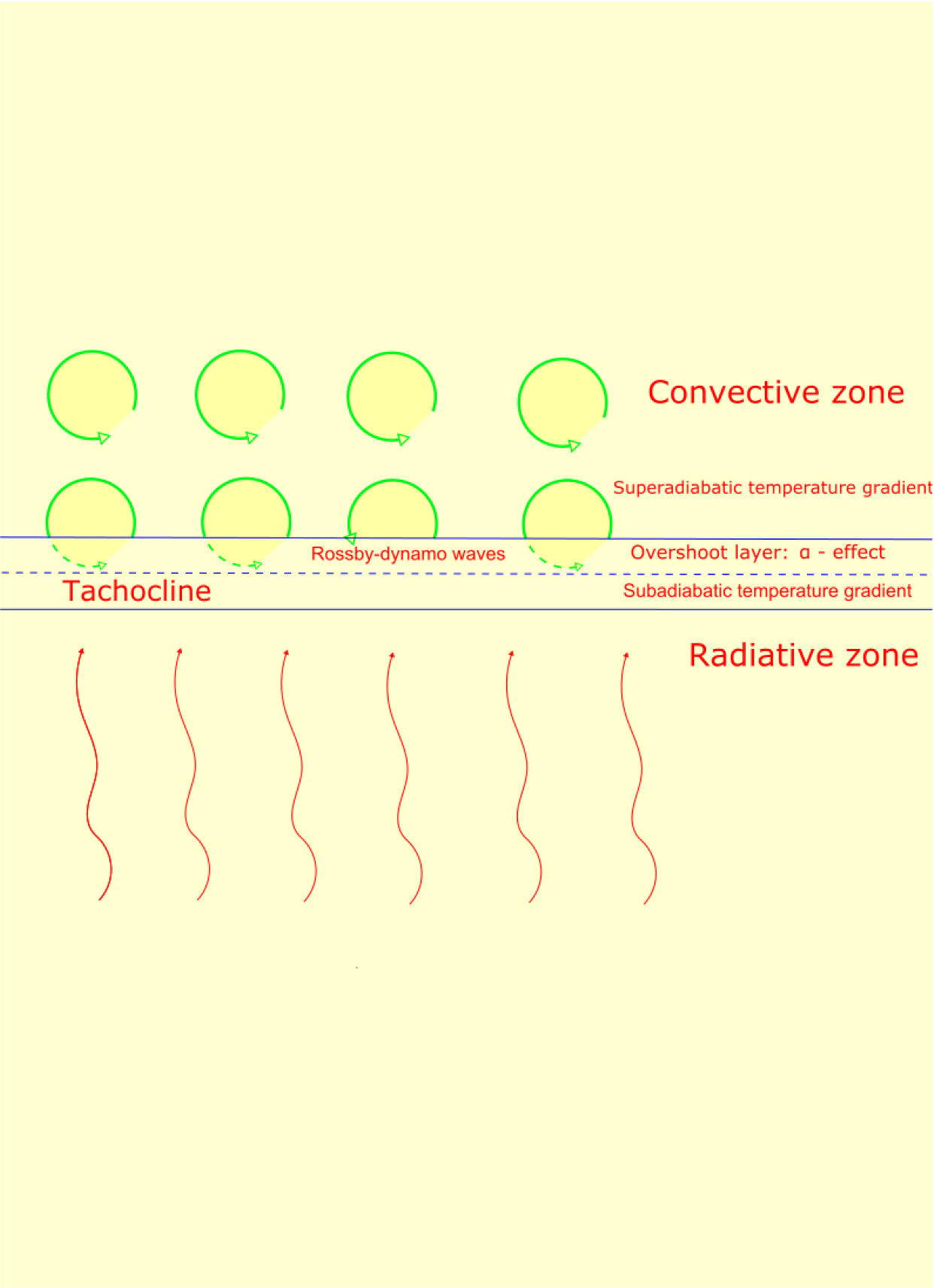}
\caption{Schematic representation of convective penetration into the tachocline, the thin interface layer between convective (upper part) and radiative (lower part) envelopes in Sun-like stars. The temperature gradient is sub-adiabatic in the tachocline and radiative envelope, but approaches  super adiabaticity  immediately above the tachocline where the convection starts. The green arrows represent the convective cells, which create the overshoot area in the upper half of the tachocline through  penetration, while the red arrows represent the radiative transport of energy by photons.  The influence of small-scale turbulent motions on the large-scale dynamics of magnetic field expressed by the $\alpha$ effect operates in the overshoot layer and causes the coupling of Rossby and dynamo waves. The image is not to scale.    \label{fig:1}}
\end{figure}

\section{MHD shallow water equations with $\alpha$ effect }

In the mean-field dynamo theory, the additional term  $\nabla {\times}\langle {\bf u}^{\prime} \times {\bf B}^{\prime} \rangle  =\nabla {\times}({\alpha {\bf B}})$ appears in the induction equation, where  ${\bf u}^{\prime} $ and ${\bf B}^{\prime} $  are the fluctuating flow and magnetic field components, the brackets $\langle  \rangle$ denote averaging,  ${\bf B}$ is the mean field, and $\alpha$ is the dynamo coefficient (Krause and R\"adler, 1980). This term is related with the mean turbulent electromotive force induced by the fluctuating flow and field components and can be rewritten as $\nabla {\times}({\alpha {\bf B}})=\nabla {\alpha} {\times}{\bf B}+{\alpha}\nabla {\times}{\bf B}$. The MHD shallow water approximation contains only the horizontal components of the induction equation as the vertical component of the magnetic field is assumed to be very small. As the horizontal component of the magnetic field does not depend on the vertical coordinate in the approximation, the second part of the dynamo term is also very small and can be neglected in the model. On the other hand, the first part of the dynamo term could be very important in the large-scale dynamics of shallow water systems. The convection is dominated in the upper overshoot part of the dynamo layer, but continuously weakens with depth, and hence the $\alpha$ coefficient depends on the vertical coordinate by definition. If one considers that the dependence is linear, as is the vertical component of the magnetic field, then the horizontal component of $\nabla {\alpha} {\times}{\bf B}$ does not depend on the vertical coordinate, and hence satisfies the shallow water approximation. Consequently, this term can be included in the induction equation of the shallow water system. 

Reduced gravity, which is an essential part of shallow water system in the tachocline, is the result of the sub-adiabatic temperature gradient providing a negative buoyancy force to the deformed upper surface \citep{Gilman2000}. Therefore, the surface is subject to a weaker gravitational field compared to the real gravity and the negative buoyancy force is proportional to $\delta$ \citep{Dikpati2001a}. We note that MHD shallow water equations with the $\alpha$ effect  are valid in the upper overshoot part of the tachocline. In the lower part of the tachocline with the stable stratification, the dynamo term in the induction equation is negligible and the reduced gravity is three orders of magnitude higher  compared to the overshoot region. Hence, here we only consider the upper overshoot part of the tachocline.

Using the new $\alpha$ term, the MHD  shallow water equations of \citet{Gilman2000} can be written in vector invariant form in the rotating system as
$$
{{\partial {\bf V}}\over {\partial t}}+{\bf \nabla} \left ( \frac{{\bf V}{\cdot}{\bf V}}{2}\right )+({\bf {\hat k}}{\times}{\bf V}){\bf {\hat k}}{\cdot}{\bf \nabla}{\times}{\bf V}-2{\bf {\hat k}}{\cdot}{\bf \Omega}({\bf V}{\times}{\bf {\hat k}})=-gH{\bf \nabla}(1+h)+
$$
\begin{equation}\label{eq1}
+{\frac{1}{4\pi \rho}}{\bf \nabla} \left ( \frac{{\bf B}{\cdot}{\bf B}}{2}\right )+{\frac{1}{4\pi \rho}}({\bf {\hat k}}{\times}{\bf B}){\bf {\hat k}}{\cdot}{\bf \nabla}{\times}{\bf B},
\end{equation}
\begin{equation}\label{eq2}
{{\partial (1+h)}\over {\partial t}}+{\bf \nabla}{\cdot}[(1+h){\bf V}]=0,
\end{equation}
\begin{equation}\label{eq3}
{{\partial {\bf B}}\over {\partial t}}= {\bf \nabla} {\times}({\bf V}{\times}{\bf B})-({\bf \nabla}{\cdot}{\bf B}){\bf V}+({\bf \nabla}{\cdot}{\bf V}){\bf B}+({\bf {\hat k}}{\cdot}{\nabla_z \alpha})({\bf B}{\times}{\bf {\hat k}}),
\end{equation}
\begin{equation}\label{eq4}
{\bf \nabla}{\cdot}[(1+h){\bf B}]=0,
\end{equation}
where ${\bf V}$ and ${\bf B}$  are the horizontal vector velocity and magnetic field, respectively, which are functions
of only the horizontal coordinates and time, ${\bf {\hat k}}$ is a unit vector in the vertical direction, ${\bf \nabla}$ is the horizontal gradient operator, ${\bf {\hat k}}{\cdot}{\bf \nabla}{\times}$  is the vertical component of the curl operator, $\bf \Omega$ is the angular frequency of rotation, $g$ is the reduced gravity, $h$ is the fractional departure of the thickness from its undisturbed value $H$, and  ${\nabla_z \alpha}$ is the vertical gradient of the dynamo coefficient, which is a function of only the horizontal coordinates. The difference between  Eqs.~(\ref{eq1})--(\ref{eq4}) and  the MHD shallow water equations of \citet{Gilman2000} is only the last term in  Eq.~(\ref{eq3}) associated with the $\alpha$ effect of the penetrative convection. Equations Eqs.~(\ref{eq1})--(\ref{eq4}) have several key differences from the equations of \citet{Deluca1986}. First,  \citet{Deluca1986} considered only the axisymmetric case, while Eqs.~(\ref{eq1})--(\ref{eq4}) contain the full horizontal extant. Next, the equations of  \citet{Deluca1986} consider only the Cartesian frame so that the rotation axis is always parallel to the vertical coordinate (called the  f-plane approximation), and therefore they do not include Rossby waves. Equations Eqs.~(\ref{eq1})--(\ref{eq4}), on the other hand, are substantially general and include the full consideration of the shallow water system. In addition, the $\alpha$ term is uniform along the vertical direction in  \citet{Deluca1986}, while here the term is a linear function of the vertical coordinate. 

The MHD shallow water equations with the  $\alpha$ effect (Eqs.~(\ref{eq1})--(\ref{eq4})) can be used to model different phenomena in the tachocline including waves, instabilities, flux transport. As an example, we studied the simplest case of its application concerning shallow water waves on the beta-plane, which also include magnetic Rossby waves modified by the $\alpha$ effect or Rossby-dynamo waves. Here we consider only the Cartesian frame on the middle latitude beta-plane, which can be generalized to a more sophisticated spherical geometry and the equatorial beta-plane.

\section{Rossby-dynamo waves in the overshoot tachocline} \label{sec:floats}

We considered the stellar overshoot tachocline as a shallow layer with uniform thickness $H$ ($\sim 10^9$ cm for the Sun) located at the distance $R$ ($\sim 5 \cdot 10^{10}$ cm for the Sun) from the stellar center \citep{Spiegel1992}. A local Cartesian frame $(x,y,z)$ on a rotating star was adopted, where $x$ is directed toward the west (i.e., in the direction of rotation), $y$ is directed toward the north, and $z$ is directed vertically outward. The layer is perceived with an unperturbed uniform toroidal magnetic field, $B_x$. We adopted solid body rotation with the angular velocity - $\Omega$ (2.9$\times$10$^{-6}$~ rad s$^{-1}$ for the sidereal angular frequency of the Sun). Differential rotation was neglected at this stage for simplicity. 

Linearized MHD shallow water equations in the rotating Cartesian frame can be written as 
\begin{equation}\label{eq5}
{{\partial u_x}\over {\partial t}}-fu_y=-gH{{\partial h}\over {\partial x}}+{\frac{B_x}{4\pi \rho}}{{\partial b_x}\over {\partial x}},
\end{equation}
\begin{equation}\label{eq6}
{{\partial u_y}\over {\partial t}}+fu_x=-gH{{\partial h}\over {\partial y}}+{\frac{B_x}{4\pi \rho}}{{\partial b_y}\over {\partial x}},
\end{equation}
\begin{equation}\label{eq7}
{{\partial b_x}\over {\partial t}}=B_x{{\partial u_x}\over {\partial x}} +{\alpha_z}b_y,
\end{equation}
\begin{equation}\label{eq8}
{{\partial b_y}\over {\partial t}}=B_x{{\partial u_y}\over {\partial x}}-{\alpha_z}b_x,
\end{equation}
\begin{equation}\label{eq9}
{{\partial h}\over {\partial t}}+{{\partial u_x}\over {\partial x}}+{{\partial u_y}\over {\partial y}}=0,
\end{equation}
where $u_x$ and $u_y$ are the velocity perturbations, $b_x$ and $b_y$ are the magnetic field perturbations, $B_x$ is the uniform unperturbed magnetic field, $h$ is the normalized perturbation of layer thickness, $g$ is the reduced gravity, ${\alpha_z}={\partial \alpha/\partial z}$ is the vertical constant gradient of the dynamo coefficient, and $f=2\Omega \sin \theta$ is the Coriolis parameter with $\theta$ being a latitude.

The coefficients of Eqs.~(\ref{eq5})--(\ref{eq9}) are not dependent on the $x$ coordinate and time, and hence can be expanded in Fourier series as $\sim \exp(-i\omega t +i k_x x)$, which leads to the single second-order equation
$$
{{d^2 u_y}\over {d y^2}}+\Bigg [{{(\omega^2-k^2_xc^2)(\omega^2-{\alpha_z}^2)-k^2_xv^2_A\omega^2}\over {c^2(\omega^2-{\alpha_z}^2)}}-
$$
\begin{equation}\label{eq10}
{{[f(\omega^2-{\alpha_z}^2)-k^2_xv^2_A{\alpha_z}]^2}\over {c^2(\omega^2-{\alpha_z}^2)(\omega^2-k^2_xv^2_A-{\alpha_z}^2)}} 
-{{k_x f{'}(\omega^2-{\alpha_z}^2)}\over {\omega(\omega^2-k^2_xv^2_A-{\alpha_z}^2)}}\Bigg ] u_y=0,
\end{equation}
where $\omega$ is the wave frequency, $k_x$ is the wave number in the toroidal direction, $c=(gH)^{1/2}$ is the surface gravity speed, and $v_A=B_x (4\pi \rho)^{-1/2}$ is the Alfv\'en speed. The prime symbol (~$'$~)  in the equation denotes the differentiation by $y$.

Hereafter we use the beta-plane approximation, which expands the Coriolis parameter in the local frame at the latitude $\theta_0$ as 
\begin{equation}\label{eq11}
f=f_0+\beta y+...,
\end{equation}
where 
\begin{equation}\label{eq12}
f_0=2\Omega\sin{\theta_0}, \,\,\, \beta={{2\Omega \cos{\theta_0}}\over {R}},
\end{equation}
and we retain only the first-order term in the expansion. Away from the equator one can assume that $\beta y \ll f_0$ and Eq.~(\ref{eq10}) can be approximated as 
$$
{{d^2 u_y}\over {d y^2}}+\Bigg [{{(\omega^2-k^2_xc^2)(\omega^2-{\alpha_z}^2)-k^2_xv^2_A\omega^2}\over {c^2(\omega^2-{\alpha_z}^2)}}-
$$
\begin{equation}\label{eq13}
{{[f_0(\omega^2-{\alpha_z}^2)-k^2_xv^2_A{\alpha_z}]^2}\over {c^2(\omega^2-{\alpha_z}^2)(\omega^2-k^2_xv^2_A-{\alpha_z}^2)}} -{{k_x \beta (\omega^2-{\alpha_z}^2)}\over {\omega(\omega^2-k^2_xv^2_A-{\alpha_z}^2)}}\Bigg ] u_y=0,
\end{equation}
where the coefficients are now independent of $y$. Consequently, the equation can be expended as $\sim \exp(ik_y y)$, which gives the dispersion equation
$$
(\omega^2-{\alpha_z}^2)[\omega (\omega^2-(k^2_x+k^2_y)c^2)(\omega^2-k^2_xv^2_A-{\alpha_z}^2) -\omega k^2_x v^2_A (\omega^2-k^2_xv^2_A)  -
$$
\begin{equation}\label{eq14}
 -(k_x \beta c^2+f^2_0 \omega)(\omega^2-{\alpha_z}^2)+2f_0{\alpha_z}k^2_xv^2_A \omega]=0.
\end{equation}

The dispersion equation has two solutions: 
\begin{equation}\label{eq15}
\omega=\pm {\alpha_z}
\end{equation}
and
$$
(\omega^2-{\alpha_z}^2)\left [\omega^3-(f^2_0+(k^2_x+k^2_y)c^2 )\omega -k_x \beta c^2 \right ]- 
$$
\begin{equation}\label{eq16}
\omega k^2_xv^2_A\left [2\omega^2-(k^2_x+k^2_y)c^2 -k^2_xv^2_A - 2f_0{\alpha_z} \right ]=0, \,\, \omega \neq \pm {\alpha_z}.
\end{equation}

The solutions of  Eq.~(\ref{eq15}) are connected to the vertical gradient of the $\alpha$ coefficient and are somewhat similar to dynamo waves \citep{Deluca1988}. Actually, the solutions are not traveling waves, but just a periodic variation of parameters including the magnetic field components, $b_x$ and $b_y$. The period of the variations depends on ${\alpha_z}={\partial \alpha/\partial z}\approx \alpha_0/l$, where $\alpha_0$ is the dynamo coefficient at the upper part of the overshoot region and $l\sim H$ is the depth of convective penetration; $H$ may take   values from 10 to 20 Mm, while the estimation of $\alpha_0$ is more complicated. Mixing length theory predicts its value near the base of convection zone as $10$ m s$^{-1}$, but numerical simulations may yield $0.1-10$ m s$^{-1}$ in order to get the solar cycle periodicity. The $\alpha$ coefficient is mostly positive in the northern hemisphere and negative in the southern hemisphere. Some numerical models of global convection also show the change in the sign near the base of the convection zone. Here we use the positive $\alpha$ in the northern hemisphere, but the dependence of wave properties on the sign must be addressed in detail in the future. Taking the width of the overshoot region as 10 Mm, the vertical gradient of the dynamo coefficient becomes $\alpha_z=\alpha_0/H\approx10^{-8}-10^{-6}$ s$^{-1}$. Then its normalized value is ${\alpha_0/(\Omega H)}\approx 0.003-0.3$. According to  Eq.~(\ref{eq15}), the corresponding period range is estimated from $\sim$ 70 days to $\sim$ 20 years. The normalized dynamo coefficient estimated from the mixing length theory, $\alpha_z/\Omega=$0.3, predicts the period of dynamo oscillations to be on the order of 100 days. In order to get the solar cycle timescale, one should take a much lower value of dynamo coefficient,  $\alpha_z/\Omega=$0.003.

The solutions of  Eq.~(\ref{eq16}) have a much richer spectrum and crucially depend on the reduced gravity speed ($c$), dynamo coefficient ($\alpha$), and Alfv\'en speed ($V_A$). \citet{Dikpati2001a} showed that the dimensionless value of reduced gravity $G=c^2/(R^2 \Omega^2)=gH/(R^2 \Omega^2)$ is proportional to $10^3 |\nabla-\nabla_{ad}|$; therefore, it is in the range of $10^{-3} \leq G \leq 10^{-1}$ in the overshoot tachocline. Then the dimensionless reduced gravity speed is $\sqrt{G}=0.03-0.3$. On the other hand, the Alfv\'en speed can be estimated as $6.3  \cdot 10^2 - 6.3 \cdot 10^4$ cm s$^{-1}$ for the magnetic field strength of 1--100 kG and the tachoclne density of 0.2 g cm$^{-3}$. Then the dimensionless value of the Alfv\'en speed, $V_A/(R\Omega)$, is in the range of $0.004 - 0.4$. 

For lower values of the Alfv\'en speed,  Eq.~(\ref{eq16}) leads to the hydrodynamic Rossby wave dispersion equation 
\begin{equation}\label{eq17}
\omega^3-(f^2_0+(k^2_x+k^2_y)c^2 )\omega -k_x \beta c^2 =0,
\end{equation} 
and therefore the influence of the $\alpha$ effect on the Rossby waves can be neglected in the weak magnetic field limit.  For $G \ll 1$ and small wave numbers, $k_x R\sim 1$, this equation has the solutions of the inertia-gravity mode, 
\begin{equation}\label{eq18}
\omega \approx \pm \sqrt{f^2_0+(k^2_x+k^2_y)c^2},
\end{equation}
and the Rossby mode, 
\begin{equation}\label{eq19}
\omega \approx -{k_x \beta c^2}[f^2_0+(k^2_x+k^2_y)c^2]^{-1}.
\end{equation}
The inertia-gravity waves have the timescale of solar rotation. On the other hand, the Rossby wave frequency of long-wavelength harmonics ($k_x R \sim 1$ ) on the middle latitudes is proportional to $\omega \sim - G \Omega$, which for $G=10^{-3}-10^{-1}$ leads to the period range of 250 days--68 years. The solar cycle timescale (11 years) is obtained for $G=6\cdot 10^{-3}$, which is in the expected range of the reduced gravity in the overshoot layer. Therefore, Rossby waves may lead to the timescales of solar cycles in reasonable agreement with the reduced gravity in the overshoot region \citep{zaqarashvili2018}. 

\subsection{MHD shallow water waves without the $\alpha$ effect}

In order to study the influence of penetrative convection on MHD shallow water waves, we first describe the waves without the $\alpha$ effect. In this case, the dispersion equation ~(\ref{eq16})  leads to the equation 
$$
\omega^4 - \left [f^2_0+(k^2_x+k^2_y)c^2 + 2k^2_xv^2_A \right ]\omega^2 -k_x \beta c^2  \omega+ 
$$
\begin{equation}\label{eq20}
k^2_xv^2_A\left [k^2_xv^2_A+(k^2_x+k^2_y)c^2 \right ]=0,
\end{equation}
which is identical to Eq. (15) of Zaqarashvili et al. (2007). This equation contains the inertia-gravity waves expressed by  Eq.  ~(\ref{eq18}) and the magneto-Rossby waves expressed by the following equation:  
\begin{equation}\label{eq21}
\omega^2 + \frac{k_x \beta c^2}{f^2_0+(k^2_x+k^2_y)c^2}  \omega - \frac{k^2_xv^2_A\left [k^2_xv^2_A+(k^2_x+k^2_y)c^2 \right ]}{f^2_0+(k^2_x+k^2_y)c^2}=0.
\end{equation}
We note that this equation is obtained when the Alfv\'en speed is much lower than the surface rotation speed. Consequently, fast ($\omega_{+}$) and slow ($\omega_{-}$) magneto-Rossby waves are governed by 
$$
\omega_{\pm} = - \frac{k_x \beta c^2}{2\left [f^2_0+(k^2_x+k^2_y)c^2\right ]} \Bigg [ 1\mp 
$$
\begin{equation}\label{eq22}
\mp \sqrt{1+\frac{4v^2_A\left [f^2_0+(k^2_x+k^2_y)c^2\right ]}{ \beta^2 c^4}\left [k^2_xv^2_A+(k^2_x+k^2_y)c^2 \right ]} \Bigg ].
\end{equation}

In the case of strong magnetic field, i.e., $v_A^2 \gg c^2$, we get 
\begin{equation}\label{eq23}
\omega_{\pm} = \pm \frac{k^2_x  v_A^2}{f_0},
\end{equation}
so the two waves have similar phase speeds. The period of the waves greatly depends on the magnetic field strength. For example, the period of  long-wavelength harmonics ($k_x R \sim 1$ ) on the middle latitudes for the field strength of 100 kG can be estimated as 156 days. In the case of weak field strength $c^2 \gg v_A^2$, we have 
\begin{equation}\label{eq24}
\omega_{+} = - \frac{k_x \beta c^2}{f^2_0} \left [ 1+ \frac{v^2_Af^2_0}{ \beta^2 c^2}(k^2_x+k^2_y)\right ]
\end{equation}
for fast magneto-Rossby waves and 
\begin{equation}\label{eq25}
\omega_{-} = \frac{k_x v^2_A}{\beta} (k^2_x+k^2_y)
\end{equation}
for the slow magneto-Rossby waves.  The period of long-wavelength harmonics ($k_x R \sim 1$ ) of fast magneto Rossby waves for 10 kG and $G=10^{-1}$ is $\sim$ 360 days and that of slow magneto-Rossby waves is $\sim$ 40 yr. The  summary of MHD shallow water waves  without the $\alpha$ effect is as follows. Lower  ($G\sim 0.01-0.001$) and higher  ($G\sim 0.1$) values of reduced gravity lead to the Schwabe (11 years) and Rieger (150-170 days) timescales, respectively, for the fast magneto-Rossby waves. The strong magnetic field (100 kG) yields the timescale of several hundred days for both fast and slow magneto-Rossby waves. On the other hand, the weaker field of 10 kG gives a significant difference in periods of the two modes of hundreds of days and several tens of years.   

\subsection{MHD shallow water waves with the $\alpha$ effect}

Now we consider the MHD shallow water waves with the $\alpha$ effect, which requires that  Eq.~(\ref{eq16}) be solved with the $\alpha$ term. First we consider the weak field limit, which results in the  equation 
\begin{equation}\label{eq26}
(\omega^2-\alpha_z^2)(\omega^3-(f^2_0+(k^2_x+k^2_y)c^2 )\omega -k_x \beta c^2) =0.
\end{equation}
The equation shows that the weak field limit decouples hydrodynamic Rossby and dynamo waves. Hence, for the weak field the $\alpha$ effect has negligible influence on Rossby waves as discussed above.

The strong field limit, i.e., when $c^2\ll v_A^2$, leads to the three types of solutions in Eq.~(\ref{eq16}). Two solutions are expressed by the equation 
\begin{equation}\label{eq27}
\omega^2_{\pm} = \frac{1}{2}\left (f_0^2+\alpha_z^2+2k_x^2v_A^2 \pm (f_0^2-\alpha_z^2)\sqrt{1+\frac{4k_x^2v_A^2}{(f_0+\alpha_z)^2}}\right ),
\end{equation}
which results in the approximate dispersion relations for high and low frequency waves as
\begin{equation}\label{eq28}
\omega_{+} \approx \pm f_0 \sqrt {1+\frac{2k_x^2 v^2_A}{f_0(f_0+\alpha_z)}}
\end{equation}
and
\begin{equation}\label{eq29}
\omega_{-} \approx \pm \alpha_z\sqrt {1+\frac{2k^2_x v^2_A}{\alpha_z(f_0+\alpha_z)}},
\end{equation}
respectively. The higher frequency waves are the inertia-gravity waves with the timescale of solar rotation, while the lower frequency waves are Rossby-dynamo waves, whose timescale depends on the value of $\alpha$ coefficient. For the value of the coefficient expected from mixing length theory, i.e., $\alpha_0 \sim$ 10$^3$ cm s$^{-1}$, we get the timescale of $\sim$ 100 days, while for the smaller coefficient of $\alpha_0 \sim$ 10 cm  s$^{-1}$, we get the timescale of years approaching  the solar cycle lengths.

\begin{figure}
   \centering
   \includegraphics[angle=0,width=10cm]{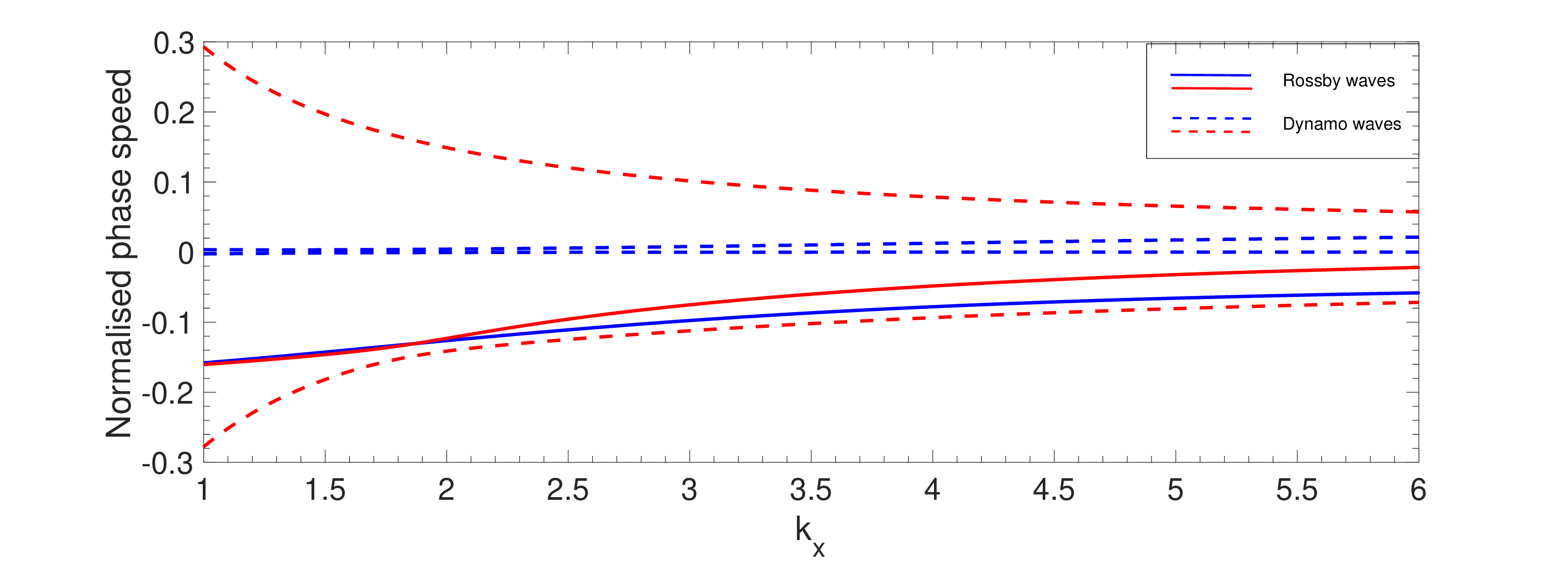}
   \includegraphics[angle=0,width=10cm]{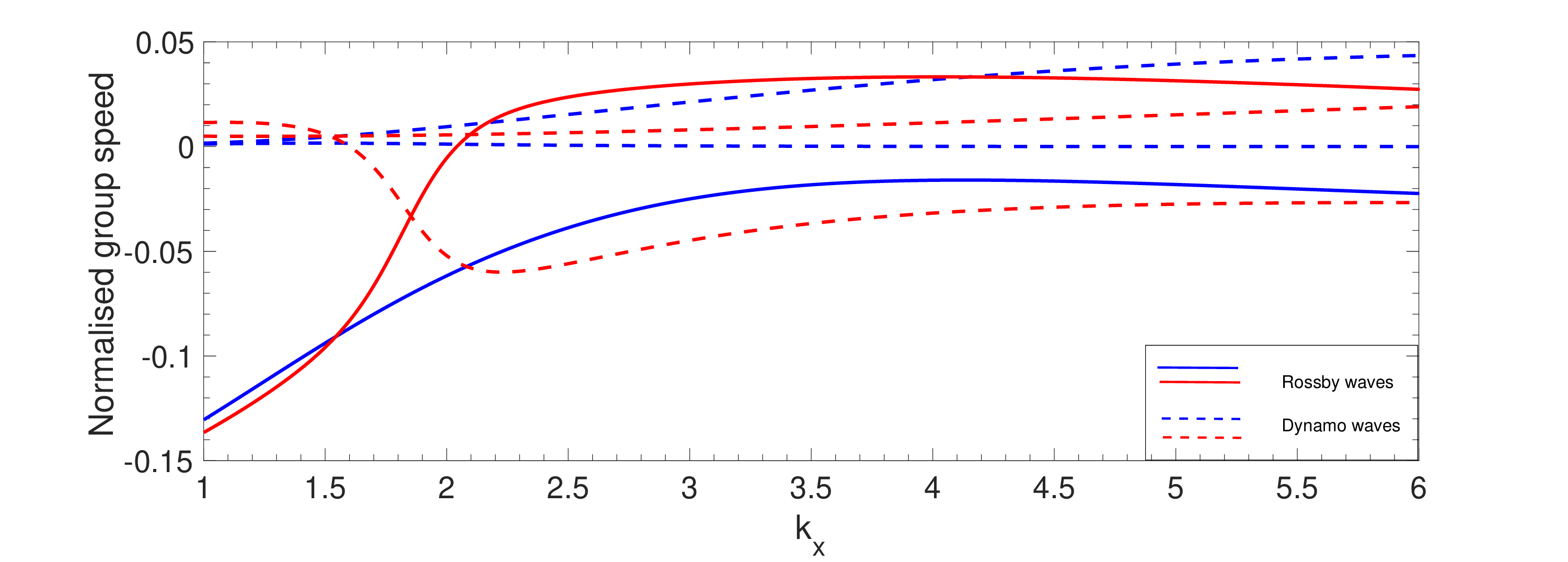}
      \caption{Phase ($\omega/k_x$, upper panel) and group ($\partial \omega/{\partial k_x}$, lower panel) speeds of Rossby-dynamo waves vs toroidal wavenumber for the toroidal magnetic field strength of 10 kG at the latitude 30$^{0}$ of the solar tachocline according to Eq. (16). The negative (positive) speeds correspond to the retrograde (prograde) propagation. The speeds are normalized by the tachocline rotation speed $\Omega R$, where $\Omega=2.9 \times$10$^{-6}$~ rad s$^{-1}$ is the sidereal angular velocity and $R \sim 5 \times 10^{10}$ cm is the distance from the solar center. The toroidal wavenumber $k_x$ is normalized by $R$. Blue and red lines indicate the phase and group speeds for the dynamo coefficient, $\alpha_z$ (normalized by $\Omega$), as 0.003 and  0.3, respectively. The dimensionless value of the surface gravity speed, $c/\Omega R=\sqrt{gH}/\Omega R$, where $g$ is the reduced gravity and $H=10^7$ m is the thickness of the overshoot layer, here is equal to 0.33. The reduced gravity is related with the fractional deference between actual and adiabatic temperature gradients, $\delta$, as  $gH/\Omega^2 R^2\sim 10^{3} \delta \sim 10^{3} |\nabla-\nabla_{ad}|$, where $\nabla=d{\ln}T/d{\ln}P$ and $\nabla_{ad}=\left ({\partial \ln T}/{\partial \ln P} \right )_{ad}$. The high frequency inertia gravity waves are not shown in these plots. We note that the group speeds of Rossby and dynamo waves reverse the propagation for higher value of dynamo coefficient (red lines): one of the dynamo waves changes from prograde to retrograde propagation at $k_x \sim 1.5$ and the Rossby wave changes from the retrograde to the prograde propagation at $k_x \sim 2$. Here $k_y=0$ is taken during computation.
              }
         \label{Figure2}
   \end{figure}

The third solution is governed by
\begin{equation}\label{eq30}
\omega \approx - \frac{ \alpha_z k_x \beta c^2}{f_0( \alpha_z f_0+2k_x^2 v^2_A)},
\end{equation}
which actually describes the Rossby wave modified by the $\alpha$ coefficient, and it gives   timescales similar to those of the hydrodynamic Rossby waves for different values of reduced gravity.

The numerical solution of the dispersion equation Eq.~(\ref{eq16}) resembles the analytical results: Rossby-dynamo waves have two retrograde and one prograde solution (upper panel of Fig. 2). For smaller and larger values of $\alpha_z$, Rossby and dynamo waves have significantly different timescales. When the dynamo coefficient has the value corresponding to that expected from the mixing length theory of the Sun ($\alpha_0=10^{3}$ cm s$^{-1}$ leading to $\alpha_z/\Omega$=0.3), then the dynamo waves (one prograde and one retrograde mode) have higher phase speeds, but for the smaller dynamo coefficient  ($\alpha_z/\Omega=0.003$), the phase speeds become very low. The phase speeds of Rossby waves do not significantly depend on the value of the dynamo coefficient (solid lines on Fig 2). The group speeds of the waves have an interesting behavior. For small dynamo coefficients, two dynamo waves have prograde groups speeds, while the Rossby wave has the retrograde group speed. For the large dynamo coefficient, the Rossby wave reverses the sign of the group speed at $k_x \sim 2$ so that the energy of the large-scale waves (wavelength $< \pi R$) propagates opposite to rotation, while the energy of shorter-scale waves propagates along the rotation. In a similar way, the group speed of one dynamo wave being positive for $k_x < 1.5$ becomes negative for  $k_x > 1.5$.

\begin{figure}
   \centering
   \includegraphics[angle=0,width=10cm]{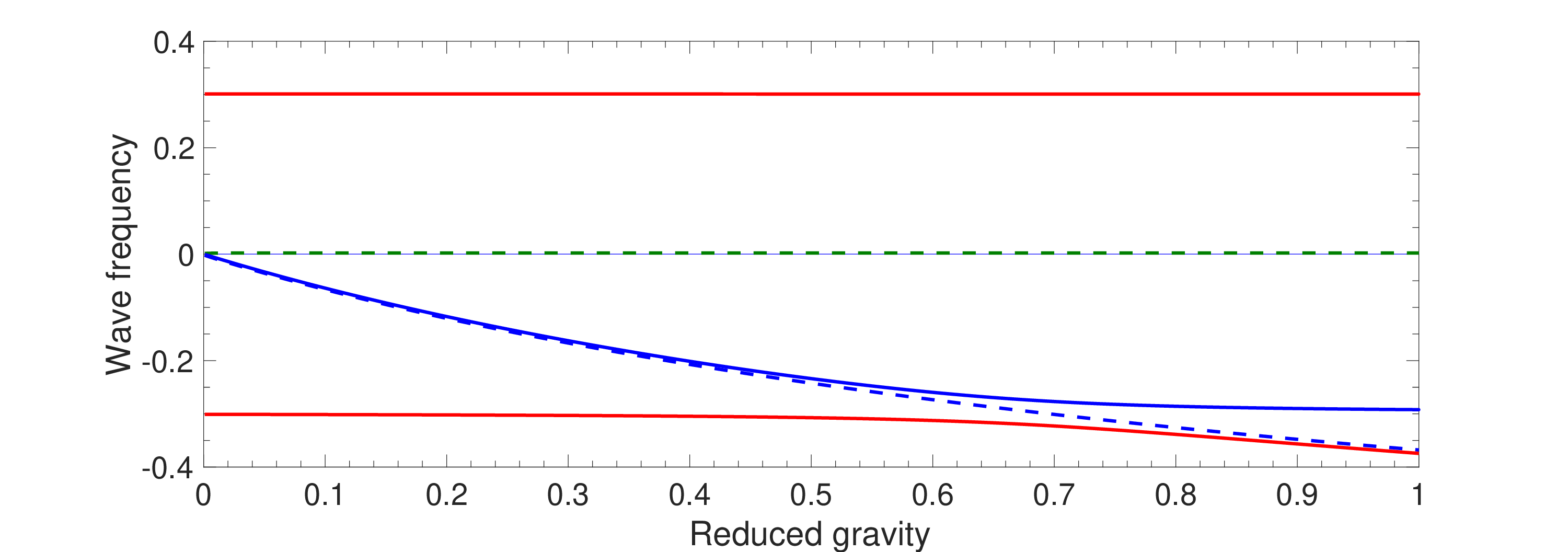}
      \caption{Wave frequency (normalized by rotation frequency) vs dimensionless reduced gravity, 
$gH/\Omega^2 R^2$ computed from Eq. (16) at the latitude 45$^{0}$ of the solar tachocline for the magnetic field strength of 10 kG and the wavelength of $k_x R=1$.  The dashed lines correspond to the solutions with $\alpha_z/\Omega=$ 0, while the solid lines show the solutions for the dynamo coefficient of $\alpha_z/\Omega=$ 0.3.  The blue and green lines show the fast and slow magneto-Rossby waves, respectively.  The red lines correspond to the modified dynamo waves. The Rossby waves have very low frequency for the lower value of reduced gravity. The frequency of slow magneto-Rossby waves remains low, but that of fast magneto-Rossby waves increases for the larger value of reduced gravity. When approaching the dynamo wave solution (for the reduced gravity of $\sim$ 0.5), the two waves do not cross, but rather switch  properties (called the avoided crossing). High frequency inertia gravity waves are not shown in the plot.  
              }
         \label{Figure3}
   \end{figure}

The frequency of Rossby-dynamo waves significantly depends on the value of reduced gravity, i.e., on the fractional deference between actual and adiabatic temperature gradients in the overshoot region of the tachocline (see Fig. 3).  When the fractional difference approaches zero, the frequencies of fast and slow magneto-Rossby waves become very low with periods of tens of years. At the same time, the frequency of dynamo waves mainly depends on the dynamo coefficient $\alpha_z$, increasing for the larger values. The prograde slow magneto-Rossby wave disappears when the dynamo coefficient increases, giving way to the  prograde dynamo wave. On the other hand, the retrograde fast magneto-Rossby waves couple with the dynamo waves for certain parameters of the dynamo coefficient and  reduced gravity. The frequencies of the two waves approach each other and exchange their dispersion properties avoiding the crossing. This happens when
\begin{equation}\label{eq}
k_x R \approx \frac{\alpha_z}{\Omega}\frac{\Omega^2 R^2}{g H}\frac{2\sin^2 \theta}{\cos \theta}.
\end{equation}
If one considers the middle latitudes and the first harmonics in toroidal direction, i.e., $k_x R \approx 1$, then  the dynamo and the Rossby waves couple on the condition that ${\alpha_z}/{\Omega} \sim gH/{\Omega^2 R^2}$. In this case, the waves can mutually exchange  energies, which means that the energy of penetrating convection (accumulated in dynamo waves) could be transferred to Rossby waves and vice versa.  

\section{Discussion}

The penetration of convection into the solar tachocline leads to the overshooting area in its upper part, which eventually leads to the mean turbulent electromotive force expressed by the $\alpha$ term in the induction equation. This is the main term in various dynamo models, which may result in the generation of a periodically reversing magnetic field explaining the cyclical behavior of solar large-scale magnetic elements.  On the other hand, the large-scale dynamics of the solar tachocline can be modeled by the shallow water approximation, which includes the behavior of the Rossby waves in the dynamo layer  \citep{Zaqarashvili2021}. Therefore, the inclusion of the dynamo term in the shallow water system results in the coupling of Rossby and dynamo waves, which may lead to  new insights into the large-scale dynamics of the solar magnetic field. The shallow water approximation considers the horizontal components of velocity and the magnetic field to be independent on the vertical coordinate, while the vertical components are linear functions of the coordinate. As the penetrative convection decreases  with depth of the tachocline, the resulting $\alpha$ effect can be also considered as the linear function of the vertical coordinate, and hence its inclusion in the shallow water approximation is possible. 

The last term of Eq. (3) corresponds to the mean electromotive force due to the penetrative convection, where  ${\nabla_z \alpha}$ is the vertical gradient of the dynamo coefficient, which is a function of only the horizontal coordinates. This term leads to the inclusion of dynamo action in the shallow water equation and to the modification of various MHD shallow water waves. First of all, the dynamo term leads to the excitation of new $\alpha$ modes in the wave spectrum of MHD shallow water system, which are  described by Eq. (15) as $\omega=\pm \alpha_z$. The frequency does not depend on wave number, and therefore the patterns are the oscillations of the magnetic field components $b_x$ and $b_y$ in time rather than  propagating waves. The timescale of the oscillations depends on the dynamo coefficient at the base of convection zone, $\alpha_0$, and the scale of convective penetration into the tachocline. For the dynamo coefficient estimated from the mixing length theory, $\alpha_0=10^3$ cm s$^{-1}$, and the convective penetration of $10$ Mm, the timescale of oscillations is $\sim$ 70 days, which for smaller dynamo coefficient of $\alpha_0=10$ cm s$^{-1}$ reaches 10--20 yr. 

On the other hand, Eq. (16) describes the MHD shallow water waves modified by the $\alpha$ effect or Rossby-dynamo waves. For the strong dynamo coefficient and weak toroidal magnetic field, the dynamo and fast magneto-Rossby waves are almost decoupled, while the slow magneto-Rossby waves disappear from the spectrum. When the dynamo coefficient decreases and the magnetic field strength increases the dynamo and Rossby waves become coupled; the modes having clear properties of magneto-Rossby and dynamo waves for the weaker field mutually change properties for the stronger field (see, e.g., Figure 3). The behavior of Rossby-dynamo waves significantly depends on the nondimensional reduced gravity, $G$.

\begin{figure}
   \centering
   \includegraphics[angle=0,width=10cm]{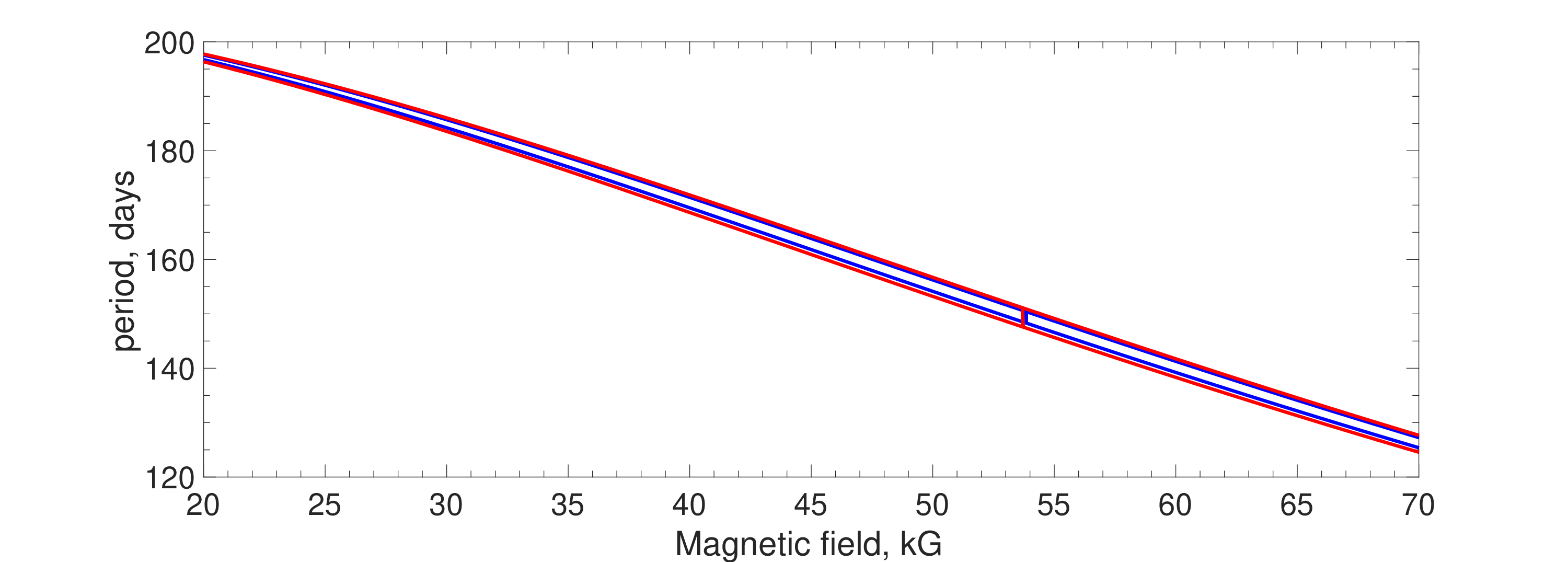}
   \includegraphics[angle=0,width=10cm]{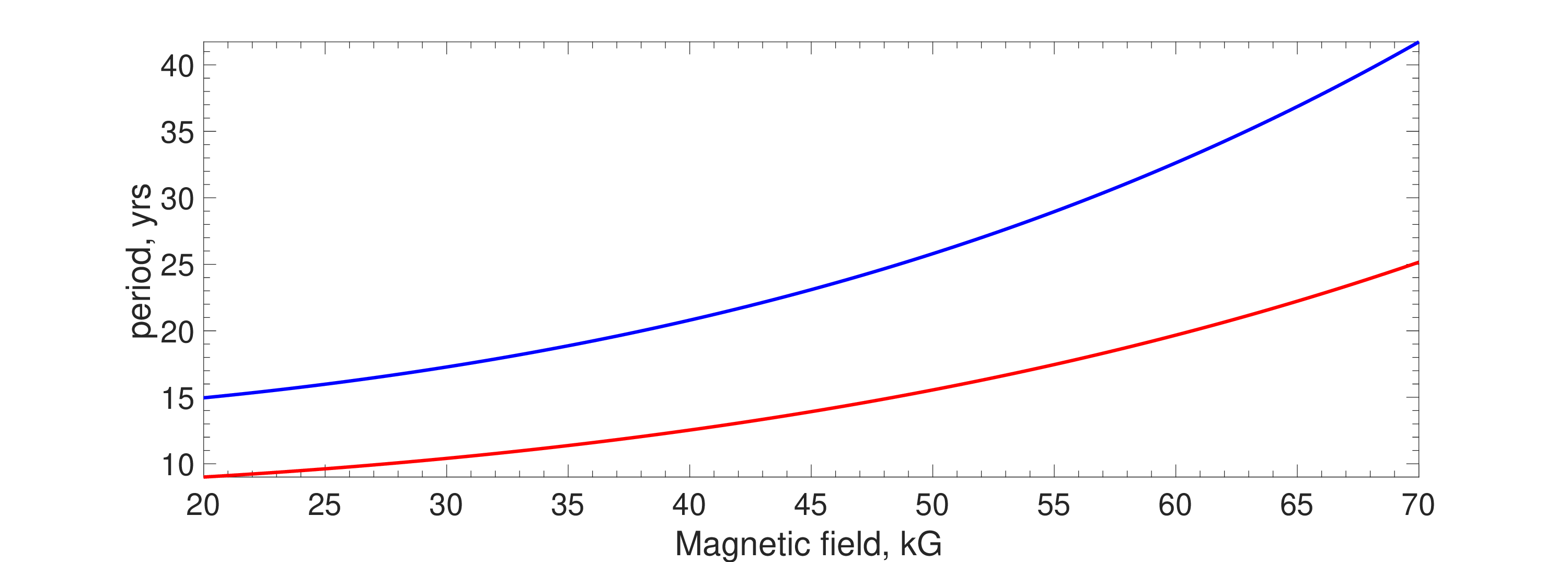}
      \caption{Period of Rossby-dynamo waves vs magnetic field strength for the dynamo coefficient, $\alpha_z/\Omega$, of 0.12  and  $k_x R=1$ computed from Eq. (16) at the latitude 30$^{0}$ of the solar tachocline. The blue and red lines correspond to the reduced gravity of $gH/\Omega^2 R^2=0.003$ and $gH/\Omega^2 R^2=0.005$, respectively.   
 Upper panel: Modified dynamo waves in the magnetic field interval of  20--70 kG, which have the Rieger cycle timescales of 130--200 days. Lower panel: Modified fast magneto-Rossby waves in the same interval of magnetic field strength, which have   solar cycle timescales of 10--40 years. 
              }
         \label{Figure4}
   \end{figure}

Solar activity variations occur over two main timescales: the solar cycle timescale of 10--20 years \citep{Schwabe1844} and the Rieger-type cycle timescales of 150--250 days \citep{Rieger1984,Carbonell1990,Oliver1998,zaqarashvili2010}. The timescale of magneto-Rossby waves significantly depends on the value of reduced gravity \citep{zaqarashvili2018}. Therefore, the different modes of Rossby-dynamo waves may correspond to the observed timescales for different parameters of dynamo coefficient and reduced gravity. The simultaneous Schwabe and Rieger timescales of Rossby-dynamo waves may arise in two different situations: 1) the high value of  $\alpha_z$ and the low value of reduced gravity, 2) the lower value of $\alpha_z$ and the higher value of reduced gravity. Figure 4 displays the periods of Rossby-dynamo waves versus a magnetic field strength of 20-70 kG in the first situation. The period of modified dynamo waves changes from 200 days at 20 kG to 130 days at 70 kG. Hence, the waves could be responsible for the occurrence of Rieger-type periodicity. The periods of the waves are almost the same for two different values of different reduced gravity. At the same time, the period of modified fast magneto-Rossby waves changes from 10-15 years to 25-40 years along the interval of 20-70 kG, and consequently could be responsible for the occurrence of Schwabe cycles. Considering the Rieger period in the interval of 160-170 days and the period of the Schwabe cycle as 22 years (full period of magnetic cycle), one can estimate the corresponding magnetic field strength as 40-45 kG for the value of $G=0.003$. The second situation is shown in Figure 5, which displays the periods of Rossby-dynamo waves versus magnetic field strengths of 0-20 kG for the small dynamo coefficient and relatively high value of reduced gravity. The period of one dynamo mode  increases for stronger magnetic fields, while the period of the second mode  decreases. The period of the Schwabe cycle (22 years) arises from the second mode at the magnetic field strength of 14 kG, while the first mode resembles the timescale of Gleissberg-type cycle \citep{Gleissberg1939}. At the same magnetic field strength, the modified fast magneto-Rossby waves lead to the Rieger-type timescales of 165-185 days.

\begin{figure}
   \centering
   \includegraphics[angle=0,width=10cm]{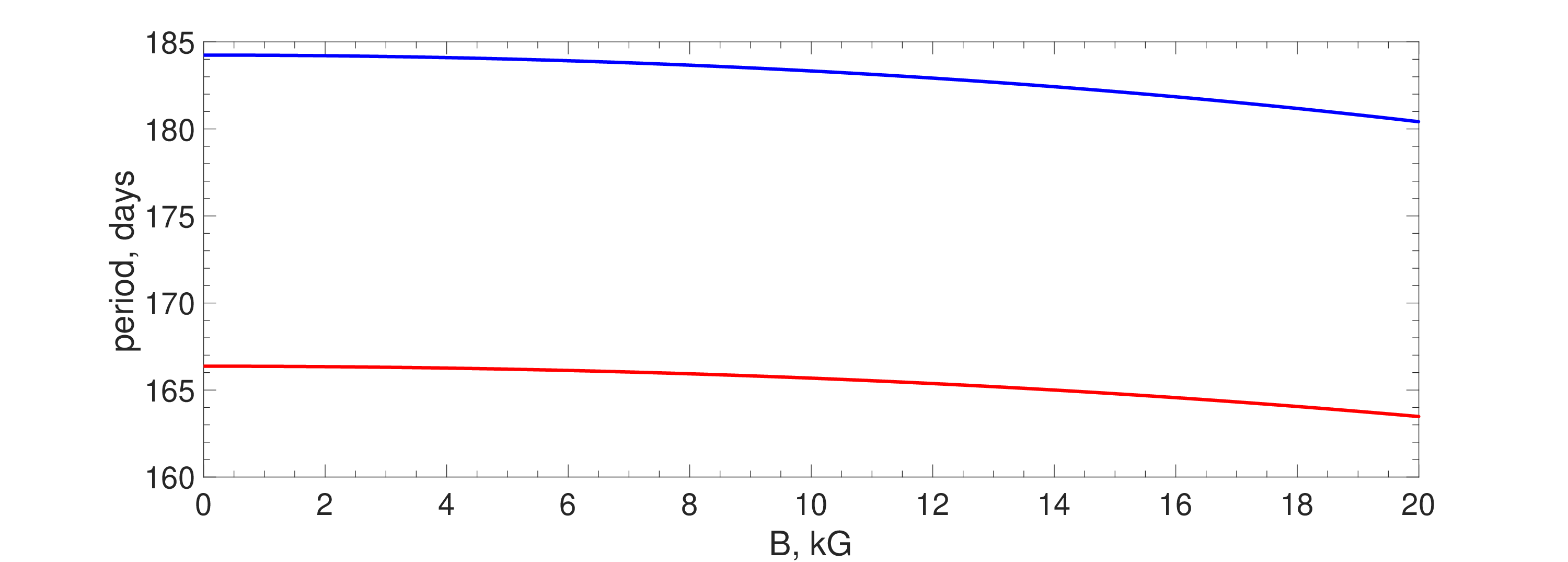}
   \includegraphics[angle=0,width=10cm]{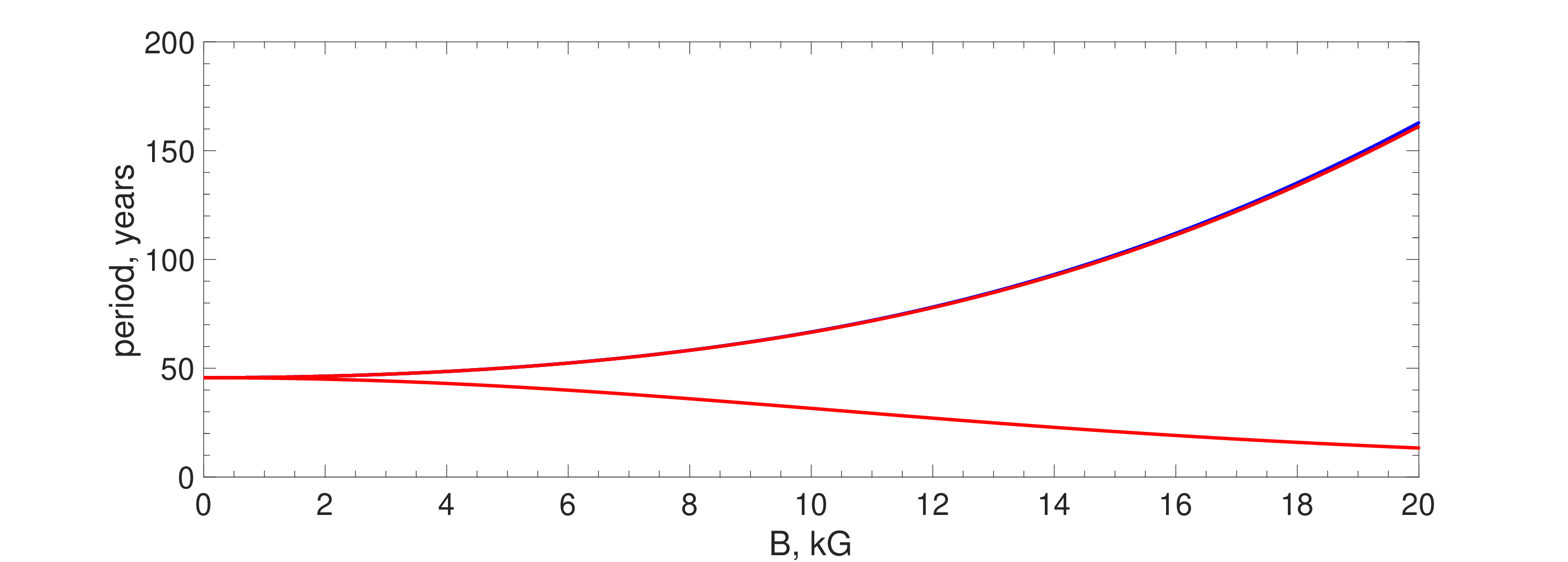}
      \caption{Period of Rossby-dynamo waves vs magnetic field strength for the dynamo coefficient, $\alpha_z/\Omega$, of 0.0015  and  $k_x R=1$ computed from Eq. (16) at the latitude 30$^{0}$ of the solar tachocline. The blue and red lines correspond to the reduced gravity of $gH/\Omega^2 R^2=0.085$ and $gH/\Omega^2 R^2=0.095$, respectively.   Lower panel: Modified dynamo waves in the magnetic field interval of  0-20 kG, on a solar cycle timescale of 15-45 years. The two solutions of modified dynamo waves correspond to a solar cycle timescale of 15-45 years (lower solution) and to Gleisberg cycle timescales of 45-160 years.  Upper panel: Modified fast magneto-Rossby waves in the same interval of magnetic field strength on  Rieger cycle timescales of 165-185 days. 
              }
         \label{Figure5}
   \end{figure}

 A detailed theory of the solar--stellar magnetic field generation and its variations requires solving a very complex astrophysical problem involving the dynamo \citep{Brun2017}. Different types of dynamos depend significantly on the toroidal field strength in the dynamo layer of  Sun-like stars. Therefore, an estimation of the dynamo field strength seems to be crucial for testing  different models. The field strength can be easily measured on the solar surface and with some caution on stellar surfaces \citep{Reiners2022}, but it is an almost impossible task for the interiors. Observed short-term cycles  \citep{Metcalfe2007, Mathur2014, Gurgenashvili2022, Breton2024} and the theory of Rossby waves \citep{Zaqarashvili2021} may lead to the estimation of magnetic field strength in internal dynamo layers of Sun-like stars \citep{Gurgenashvili2016}. This may become an important tool for the sounding of stellar interiors alongside with asteroseismology. Rossby-dynamo waves obtained from the shallow water equations with the penetrative convection will enrich the wave spectrum in tachoclines and can play invaluable role in the magneto-seismology of solar--stellar interiors. The waves also operate for a weak seed magnetic field and can generate periodic toroidal and poloidal components resembling solar magnetic field properties. In certain parameters of reduced gravity and overshooting, the timescale of periodic reversals agrees with the solar cycle period showing the importance of the proposed mechanism for the dynamo and magnetic activity.  The new paradigm of magnetic field generation by Rossby-dynamo waves will drive breakthrough research  in the study of Sun-like stars.

\section{Conclusion}

The large-scale influence of overshooting convection on the upper part of the tachocline was added to the MHD shallow water equations of Gilman (2000) as a separate term with the $\alpha$ effect in the induction equation.  A simple linear application of the equations to the study of MHD waves in the mid latitude $\beta$-plane approximation gives an interesting behavior of magneto-Rossby and the $\alpha$ (or dynamo) waves. The wave dynamics depend on the values of reduced gravity, magnetic field strength, and the $\alpha$ parameter.  Magneto-Rossby and $\alpha$ waves are coupled in certain parameters and consequently can be mentioned as Rossby-dynamo waves. The waves may drive the oscillations on different timescales in the tachocline magnetic field, and hence in solar activity. It is shown that the weaker toroidal magnetic field strength of $\sim$ 10 kG may lead to the observed activity cycles on the Sun; modified fast magneto-Rossby and $\alpha$ waves could excite Rieger-type and Schwabe cycles, respectively, for the smaller dynamo coefficient and for the higher value of reduced gravity. On the other hand, in the case of stronger field strength of $\sim$ 50 kG, the modified fast magneto-Rossby and $\alpha$ waves could drive the Schwabe and Rieger cycles, respectively, for the larger dynamo coefficient and for the lower value of reduced gravity. Further application of the formalism to the internal dynamics (also with numerical simulations) is an important research direction in solar and stellar physics.

\begin{acknowledgements}
This work was supported by the Austrian Science Fund (FWF) project PAT7550024 and by Shota Rustaveli National Science Foundation of Georgia (project FR-21-467). This work is also supported by the NSF National Center for Atmospheric Research, which is a major facility sponsored by the National Science Foundation under cooperative agreement 1852977. MD acknowledges support from several NASA grants, namely NASA-HSR award 80NSSC21K1676, Stanford COFFIES Phase II NASA-DRIVE Center subaward 80NSSC22M0162, and NASA-HSR subaward 80NSSC21K1678 from JHU/APL. This research was supported by the International Space Science Institute (ISSI) in Bern, through ISSI International Team project 24-629 (Multi-scale variability in solar and stellar magnetic cycles). We thank Ed Deluca for his thorough review and for helpful comments, which have significantly improved our paper.
\end{acknowledgements}

\bibliographystyle{aa} 
\bibliography{aa56108-25}

\end{document}